\documentstyle[12pt,psfig,epsf]{article}
\newcommand{\qua}{(\gamma^{m}D_{m})^{2}}

\newcommand{\eff}{E_{eff(2+1)}}
\newcommand{\effp}{E_{eff(3+1)}}
\newcommand{\del}{\delta_{l,s}(k,r)}
\newcommand{\opa}{\frac{1}{2}tr\ln(\not\!\! D^{2}+m^{2}_{f})}
\begin{document}

\title{ Fermion-induced effective action in the presence of a static inhomogeneous magnetic field}
\author{Pavlos Pasipoularides \footnote{paul@central.ntua.gr}\\
       Department of Physics, National Technical University of
       Athens \\ Zografou Campus, 157 80 Athens, Greece}
\date{ }
       \maketitle

\begin{abstract}
 We present a numerical study of the fermion-induced
 effective action in the presence of a static inhomogeneous magnetic field for
 both $3+1$ and $2+1$ dimensional QED using a novel approach. This
 approach is appropriate for cylindrically symmetric magnetic
 fields with finite magnetic flux $\Phi$. We consider families
 of magnetic fields, dependent on two parameters, a typical
 value $B_{m}$ for the field and a typical range $d$.
 We investigate the behavior of the effective action for
 three distinct cases: 1) keeping $\Phi$ (or $B_{m}d^{2}$) constant and varying
 $d$, 2) keeping $B_{m}$ constant and varying $d$
 and 3) keeping $d$ constant and varying $\Phi$ (or $B_{m}d^{2}$).
 We note an interesting difference in the limit
 $d\rightarrow +\infty$ (case 2) between smooth and discontinuous
 magnetic fields. In the strong field limit (case 3) we also derive
 an explicit asymptotic formula for the $3+1$ dimensional action.
 We study the stability of the magnetic field and we show
 that magnetic fields of the type we examine remain unstable, even
 in the presence of the fermions.
 In the appropriate regions we check our numerical results against the
 Schwinger formula (constant magnetic field), the derivative expansion
 and the numerical work of M. Bordag and K. Kirsten.
 The role of the Landau levels for the effective action, and
 the appearance of metastable states for large magnetic flux,
 are discussed in an appendix.
\end{abstract}

\section{Introduction}
The evaluation of the fermion-induced effective action in the presence of a static
magnetic field is important in quantum electrodynamics. Historically, the effective
action for a homogeneous magnetic field was first computed by Heisenberg and Euler
\cite{1}, then by Weisskopf \cite{2}, and later by Schwinger \cite{3} with the proper
time method. The effective action in this case can be expressed by an explicit
formula. In reference \cite{4} Redlich obtained the corresponding results for the
$2+1$ dimensional QED.

For inhomogeneous magnetic fields explicit formulas are not available. Approximations
can be developed for the case of $a)$ weak magnetic fields (perturbative expansion)
and $b)$ smooth magnetic fields (derivative expansion).

More recently, for a special form of the magnetic field it has been possible to derive
a closed formula for the effective action for both $3+1$ and $2+1$ dimensions
\cite{5,9}. Another exact result is presented in reference \cite{del}, for a magnetic
field with a delta function profile in two dimensional Euclidean space.

The aim of this paper is to obtain, by $\mathit{numerical}$ computations, qualitative
knowledge of the dependence of the fermion-induced effective action on various
quantities characterizing the magnetic field such as its size and its degree of
inhomogenity etc. For this numerical study we use a simple novel approach for the fast
computation of the effective action, which is appropriate for static, cylindrically
symmetric magnetic fields with finite magnetic flux $\Phi$. We will call such fields
magnetic flux tubes.

Previous works towards the numerical study of the effective action
are the following: in reference \cite{10} the effective energy, in
$3+1$ dimensions, is computed numerically for a cylindrically
symmetric static magnetic field which is constant inside a
cylinder and zero outside it. In a more recent work \cite{11} the
effective action was computed for a cylindrically symmetric static
magnetic field with a delta function profile, in 3+1 dimensions.
Note that this field has infinite classical energy.

Our purpose in this paper is to go beyond the previous numerical
works, and to study, for the first time, the $2+1$ QED effective
energy in the presence of a magnetic flux tube. In addition, in
the case of $3+1$ dimensions we consider more realistic magnetic
field configurations than those of the previous works, for example
the Gaussian magnetic field of Eq. (2) bellow.

We study families of magnetic fields dependent on two parameters,
a typical value $B_{m}$ \footnote{We define this typical magnetic
field strength as $B_{m}=\int \vec{B}\cdot d\vec{S} /\pi d^{2}$.}
for the field (or the magnetic flux $\Phi=B_{m}d^{2}/2$) and a
typical range $d$. We investigate the behavior of the effective
action for three distinct cases: 1) keeping $\Phi$ (or
$B_{m}d^{2}$) constant and varying $d$, 2) keeping $B_{m}$
constant and varying $d$ and 3) keeping $d$ constant and varying
$\Phi$ (or $B_{m}d^{2}$). The mass of the fermion $m_{f}$ is
assumed to be constant.

We note a difference in the limit $d\rightarrow +\infty$ (case 2)
between smooth and discontinuous magnetic fields (section 5). We
will see that this difference is related to the fact that the
derivative expansion fails for discontinuous magnetic fields like
the magnetic field of Eq. (3) below.

An interesting problem not covered by previous works, is the
behavior of the effective energy for large magnetic flux $\phi$
(strong magnetic field) and fixed $m_{f}d$, where $\phi=e \Phi/2
\pi$. In section $7$ we present an investigation for large $\phi$.
Especially in the case of $3+1$ dimensions, we derive an explicit
asymptotic formula for $\phi\gg 1$.

We also consider the question of stability of the magnetic field
in the presence of the fermions. The question is, whether some
radius exists that minimizes the total energy (classical energy
plus effective energy) for fixed magnetic flux $\Phi$, thus
rendering the magnetic field stable at the quantum level. The same
question was considered by the authors of reference \cite{10}, for
the magnetic field of Eq. (3) (below) in the case of $3+1$
dimensions, and the answer was negative. Our purpose here is to
study the stability of magnetic flux tubes in the case of $2+1$
dimensions. In addition in the case of $3+1$ dimensions, we
generalize the results of reference \cite{10} for more realistic
magnetic field configurations like the Gaussian magnetic field of
Eq. (3). Note, that we answer this question for several
two-parameter families of magnetic field configurations of the
type of flux tube for both 2+1 and 3+1 dimensions (see the
magnetic fields at the end of the present section).

In appendix C we explain the role of the Landau levels for the
effective action, relating them to the metastable states which
appear in the case of large magnetic flux $\Phi$ (see appendix B).

The vector potential and the magnetic field for the class of magnetic fields we
examine, are given by the following equations:
\begin{equation}
\vec{A}(r)=\frac{r}{2}\:f(r) \: \vec{e_{\phi}}, \qquad
\vec{B}(r)=\frac{1}{2r}\frac{d}{dr}(r^{2}\:f(r)) \: \vec{e_{z}}
\end{equation}
where $f(r)$ is a function with asymptotic behavior $2\phi/r^{2}$ as r tends to
infinity and $\phi=e\Phi/2\pi$. In what follows, we assume the rescaling $B
\rightarrow e B$ except where we state otherwise. We present our numerical results for
two magnetic fields, one with a Gaussian profile and one which is constant inside and
vanishes abruptly outside a cylinder of radius $d$ with respective field strengths
\begin{eqnarray}
B_{1}(r)&=&\frac{2\phi}{d^{2}}\: \exp(-\frac{r^{2}}{d^{2}})\\
B_{2}(r)&=&\frac{2\phi}{d^{2}}\: \theta(d-r)
\end{eqnarray}
where $\theta(x)$ is the step function.

We have performed analogous calculations for six two-parameter
families of magnetic fields other than those of Eqs. (2) and (3),
with magnetic field strengths
\begin{eqnarray}
&&B_{3}(r)=\frac{2\phi}{d^{2}}\;
\frac{1}{\cosh^{2}(r^{2}/d^{2})}\:, \quad
B_{4}(r)=\frac{4\phi}{d^{2}}\;\left(\frac{r^{2}}{d^{2}}\right)
\exp(-\frac{r^{4}}{d^{4}})\:,\quad
B_{5}(r)=\frac{4\phi}{d^{2}\pi}\: \frac{1}{(r^{4}/d^{4}+1)}
\nonumber \\ &&B_{6}(r)=\frac{6\phi}{d^{2}}\:(1-\frac{r}{d})\:
\theta(d-r)\:, \quad
B_{7}(r)=\frac{4\phi}{d^{2}}\:(1-\frac{r^{2}}{d^{2}})\:\theta(d-r)\:,
\quad B_{8}(r)=\frac{2\phi}{d^{2}}\: \frac{1}{(r^{2}/d^{2}+1)^{2}}
\nonumber
\end{eqnarray}
Since including the results in detail would make this paper
unnecessarily long it is perhaps sufficient to state that our
conclusions (about the stability of the magnetic field or the
dependence on $d$ of the effective action etc) for the magnetic
fields of Eqs. (2) and (3) are valid also for the above-mentioned
magnetic fields.

\section{Effective energy}
 We consider the following path integral that corresponds to QED
\begin{equation}
Z=\int {\cal D} A {\cal D}\bar{\Psi} {\cal D} \Psi e^{i\int d^{D}x
\: (-\frac{1}{4}F_{\mu\nu}F^{\mu\nu}+\bar{\Psi}(i \: \not \!
D-m_{f})\Psi)}
\end{equation}
in dimension D $(D=3,4)$. Integrating out the fermionic degrees of freedom in the
above mentioned path integral, we obtain the effective action expressed as the
logarithm of a determinant
\begin{equation}
S_{eff}[A]=\frac{1}{i}\ln\int {\cal D}\bar{\Psi} {\cal D}\Psi
e^{i\int d^{D}x\: \bar{\Psi}(i \: \not \!
D-m_{f})\Psi}=\frac{1}{i}tr\ln(i \not\!\!D-m_{f})
\end{equation}
where  $\not\!\!D= \gamma^{\mu}(\partial_{\mu}-ieA_{\mu})$, and
$\gamma^{0}=\sigma_{3}$, $  \gamma^{1}=i\sigma_{1}$, $ \gamma^{2}=i\sigma_{2}$  is a
two component representation of the gamma matrices for the $2+1$ dimensional QED. A
four component representation will be used for the 3+1 dimensional case.

We will use the identity
\begin{equation}
tr\ln(i \not\!\!D-m_{f})=\opa
\end{equation}
in order to take advantage of the diagonal form of the operator
$\not\!\!D^{2}+m^{2}_{f}$. It is well known that for $2+1$ dimensional QED a parity
violating term (Chern-Simons term) is induced \cite{4,cer}.  However, this term is
zero for the case we investigate \footnote{The induced Chern-Simons term is of the
form $\frac{\kappa}{2}\int d^{3}x \: \varepsilon^{\mu\nu\rho}A_{\mu}
\:\partial_{\nu}A_{\rho}$ and it is zero, for our case, since $A_{o}=0$ and
$\dot{A}_{1}=\dot{A}_{2}=0$.}. See also the relevant comments in reference \cite{sim}.

We deal first with the $2+1$ dimensional problem. The operator
$\not\!\!D^{2}+m^{2}_{f}$ for a static magnetic field can be put into the form
\begin{equation}
\not\!\!
D^{2}+m^{2}_{f}=\partial_{0}^{2}+(\gamma^{m}D_{m})^{2}+m^{2}_{f}\qquad
(m=1,2)
\end{equation}
This operator has a complete system of eigenfunctions of the form
$\Psi_{\{n\}}(\vec{x},t)=e^{-i\omega t}\Psi_{\{n\}}(\vec{x})$
where $\Psi_{\{n\}}(\vec{x})$ satisfies the eigenvalue equation
$(\gamma^{m}D_{m})^{2}\Psi_{\{n\}}(\vec{x})=E_{\{n\}}\Psi_{\{n\}}(\vec{x})$
and $\{n\}$ is a set of quantum numbers. Note that $E_{\{n\}}\geq
0$ since the operator $(\gamma^{m}D_{m})^{2}$ is positive definite
as the square of a hermitian operator. It is convenient here to
refer to the effective energy $E_{eff}=- S_{eff}/T$ instead of the
effective action $S_{eff}$, where $T$ is the total length of time.
The effective energy, if we perform a Wick rotation of the
integration variable $\omega$ $(\omega\longrightarrow i\omega)$,
is given by
\begin{equation}
E_{eff(2+1)}=-\frac{1}{2\pi}\sum_{\{n\}}\int_{0}^{+\infty}d\omega
\ln(\omega^{2}+E_{\{n\}}+m^{2}_{f})
\end{equation}

We make the above integral convergent by subtracting a term which
is independent of the magnetic field. We may do so because our
purpose is to compute the difference between the effective energy
in the presence of the magnetic field, and the energy when the
magnetic field is absent. If we subtract the term
$-\frac{1}{2\pi}\sum_{\{n\}}\int_{0}^{+\infty}d\omega
\ln(\omega^{2}+m^{2}_{f})$ from (8) and integrate, we get
\begin{equation}
E_{eff(2+1)}=-\frac{1}{2}\sum_{\{n\}}\left(\sqrt{E_{\{n\}}+m^{2}_{f}}-m_{f}\right)
\end{equation}
This expression incorporates the expectation that very massive fermions interact
weakly with the magnetic flux tube (i.e the effective energy tends to zero as $m_{f}$
tends to infinity).

The extension to the 3+1 dimensional case can be made if we replace $m_{f}^{2}$ by
$k_{3}^{2}+m_{f}^{2}$ in (9) and integrate over the $k_{3}$ momentum. An overall
factor of $2$ (from the Dirac trace) must also be included. For details see e.g
reference \cite{9}.
\begin{equation}
E_{eff(3+1)}=-L_{z}\sum_{\{n\}}\int_{-\infty}^{+\infty}\frac{dk_{3}}{2\pi}\left(\sqrt{k_{3}^{2}+E_{\{n\}}+m^{2}_{f}}-\sqrt{k_{3}^{2}+m^{2}_{f}}\right)
\end{equation}
where $L_{z}$ is the length of the space box in the $z$ direction.

Introducing a momentum cut-off $\Lambda/2$ we find:
\begin{eqnarray}
E_{eff(3+1)}&=&\frac{1}{4\pi}\sum_{\{n\}}(E_{\{n\}}+m_{f}^{2})\ln(\frac{E_{\{n\}}+m_{f}^{2}}{m_{f}^{2}})-\frac{1}{4\pi}\sum_{\{n\}}E_{\{n\}}\nonumber
\\& &-\frac{1}{4\pi}\ln(\frac{\Lambda^{2}}{m_{f}^{2}})
\sum_{\{n\}}E_{\{n\}}
\end{eqnarray}
where we have omitted $L_{z}$ and we assume that we calculate
energy per unit length.

For large $m_{f}$ the above expression for the effective energy
must tend to the first diagram (vacuum polarization diagram) of
the perturbative expansion of the effective energy:
\begin{eqnarray}
E_{eff(3+1)}^{(2)}=&-& \frac{1}{8\pi^{3}}\int_{0}^{1} dx x (1-x)
\int_{0}^{+\infty}dq q |\widetilde{B}(q)|^{2}
\ln(\frac{m_{f}^{2}+q^{2}x (1-x)}{m_{f}^{2}}) \nonumber \\
&+&\frac{1}{24\pi^{2}}\ln(\frac{\Lambda^{2}}{m_{f}^{2}})\int
d^{2}\vec{x} B^{2}(\vec{x})
\end{eqnarray}
where $\widetilde{B}(q)=\int
d^{2}\vec{x}\:e^{-i\vec{q}\:\cdot\vec{x}}B(\vec{x})$. Note that
$\widetilde{B}(q)$ depends only on $q=|\vec{q}|$ as we have
assumed that the magnetic field is cylindrically symmetric.

 If we compare the equations (11) and (12) for $m_{f}\rightarrow\infty$ we
find
\begin{equation}
\sum_{\{n\}}E_{\{n\}}=-\frac{1}{6\pi}\int d^{2}\vec{x}
B^{2}(\vec{x})
\end{equation}
Thus, the logarithmically divergent part in equation (11) can be
dropped if we include an appropriate counterterm in the QED
Lagrangian. This counterterm is defined by the on-shell
renormalization condition \footnote{For the definition of the
vacuum polarization function $\Pi(q^{2})$ and for details about
the renormalization condition for $\Pi(q^{2})$  see, for example,
the reference \cite{pesk}.} $\Pi(0)=0$. The renormalized effective
energy is then given by the equation
\begin{eqnarray}
E_{eff(3+1)}^{(ren)}=\frac{1}{4\pi}\sum_{\{n\}}(E_{\{n\}}+m_{f}^{2})\ln(\frac{E_{\{n\}}+m_{f}^{2}}{m_{f}^{2}})-\frac{1}{4\pi}\sum_{\{n\}}E_{\{n\}}
\end{eqnarray}

Note, that the renormalized effective energy vanishes for large values of $m_{f}$, as
expected. For the sake of simplicity we will drop the index $(ren)$ in the rest of
this paper.

In order to define the effective action for massless fermions for which the right hand
side of Eq. (14) is singular, we should impose the renormalization condition for the
vacuum polarization funtion $\Pi(q^{2})$ at a spacelike momentum $-M^{2}$
($\Pi(-M^{2})=0$). Accordingly, a formula for the effective action for massless
fermions is given in appendix A.

In cylindrical coordinates and for the special magnetic field of
Eq. (1), the operator $\qua$ has the following diagonal form
\begin{eqnarray}
(\gamma^{m}D_{m})^{2}&=&-D^{2}_{1}-D^{2}_{2}-s B \nonumber
\\ &=&-\frac{\partial^{2}}{\partial
r^{2}}-\frac{1}{r}\frac{\partial}{\partial
r}-\frac{1}{r^{2}}\frac{\partial^{2}}{\partial\phi^{2}}-\frac{1}{i}\frac{\partial}{\partial\phi}
f(r)+\frac{r^{2}f^{2}(r)}{4}-s B(r)
\end{eqnarray}
where $s$ takes the values $\pm1$ that correspond to the two possible spin states of
the electron. Because the operator $\qua$ is invariant under rotations around the $z$
axis, the eigenfunctions $\Psi_{\{n\}}(\vec{x})$ have the form
$\Psi_{\{n\}}(\vec{x})=e^{il\phi}u(r)/ \sqrt{r}$, where $l$ is the quantum number of
the angular momentum $(l=0,\pm1,\pm2,...)$. The function $u(r)$ satisfies the equation
\begin{eqnarray}
\left(-\frac{d^{2}}{dr^{2}}+\frac{l^{2}-\frac{1}{4}}{r^{2}}+v_{l,s}(r)\right)u(r)=k^{2}u(r)
\end{eqnarray}
where
\begin{equation}
v_{l,s}(r)=-l f(r)-s B(r)+\frac{r^{2}f^{2}(r)}{4}
\end{equation}
and $k^{2}$ is the corresponding eigenvalue (we may set $E_{\{n\}}=k^{2}$ since
$E_{\{n\}}\geq 0$). The spectrum of the radial Schrodinger equation (16) besides the
continuous spectrum $(0,+\infty)$, also may have zero modes according to the
Aharonov-Casher theorem \footnote{According to this theorem a zero mode exists if the
conditions $1<l+1<\phi$ and $s=1$ are satisfied.}\cite{12}. Note that for our special
case $\{n\}=\{k,l,s\}$.

\section{Density of states}
In order to sum up over the continuous modes of the equation (16), we shall need the
density of states $\frac{dn}{dk}=\rho_{l,s}(k)$ (i.e the number of states per unit
$k$).
\begin{equation}
\rho_{l,s}(k)=\rho_{l,s}^{(free)}(k)+\frac{1}{\pi}\frac{d\delta_{l,s}(k)}{dk}
\end{equation}
where $\rho_{l,s}^{(free)}(k)=\frac{\pi}{L}$ is the density of states for free space,
and $\delta_{l,s}(k)$ is the phase shift which corresponds to $l^{th}$ partial wave
with momentum $k$ and spin $s$. The relation (18) is shown e.g in reference
\cite{den}.

A noteworthy feature of the function $ \delta_{l,\sigma}(k)$ is
that it exhibits jumps which corresponds to metastable states for
large values of $\phi$. This is shown in Fig. \ref{8} of appendix
A.

If we use the relations (18) and (9), the effective energy for $2+1$ dimensional QED
is
\begin{eqnarray}
E_{eff(2+1)}=-\frac{1}{2\pi}\int_{0}^{+\infty}\left(\sqrt{k^{2}+m^{2}_{f}}-m_{f}\right)\:\frac{d}{dk}\left(\sum_{l,s}\delta_{l,s}(k)\right)\:
dk
\end{eqnarray}
where $\rho_{l,s}^{(free)}(k)$ was dropped since it contributes a field independent
term. Note that the zero modes do not contribute explicitly in (19).

The series of the phase shifts over $l$ in (19) is not absolutely convergent. The
simplest way to define it, is to sum symmetrically over $l$. We define the function
\begin{equation}
\Delta(k)=\lim_{L\rightarrow+\infty}\sum_{s,l=-L}^{L}\delta_{l,s}(k)
\end{equation}
The asymptotic behavior of the phase shifts for $|l|\gg k d $ is
\begin{equation}
\delta_{l,s}(k)\rightarrow\left(-\frac{|l-\phi|}{2}+\frac{|l|}{2}\right)\pi
\end{equation}
and the sum $ \delta_{l,s}(k)+\delta_{-l,s}(k) $ vanishes for enough large values of
$l$ so the limit in (20) exists.

The large $l$ asymptotic behavior can be obtained from the WKB approximation for the
phase shifts:
\begin{equation}
\delta_{l,s}(k)=Re\left\{\int_{0}^{+\infty}\left(\sqrt{k^{2}-v_{l,s}(r)-\frac{l^{2}}{r^{2}}}-\sqrt{k^{2}-\frac{l^{2}}{r^{2}}}\right)
dr \right\}
\end{equation}
For large $l$ only the asymptotic $O(1/r^{2})$ tail of the potential contributes to
the above integral. Replacing $v_{l,s}(r)$ by its asymptotic form
$\frac{\phi^{2}-2\phi l}{r^{2}}$ yields (21).

Note that an asymmetric sum would give the same result since
$\lim_{L\rightarrow+\infty}\sum_{s,l=-L}^{L+\Delta L}\delta_{l,s}(k)=\Delta(k)+c'$ and
$c'$ is independent of $k$.

\begin{figure}[h]
\centerline{\hbox{\psfig{figure=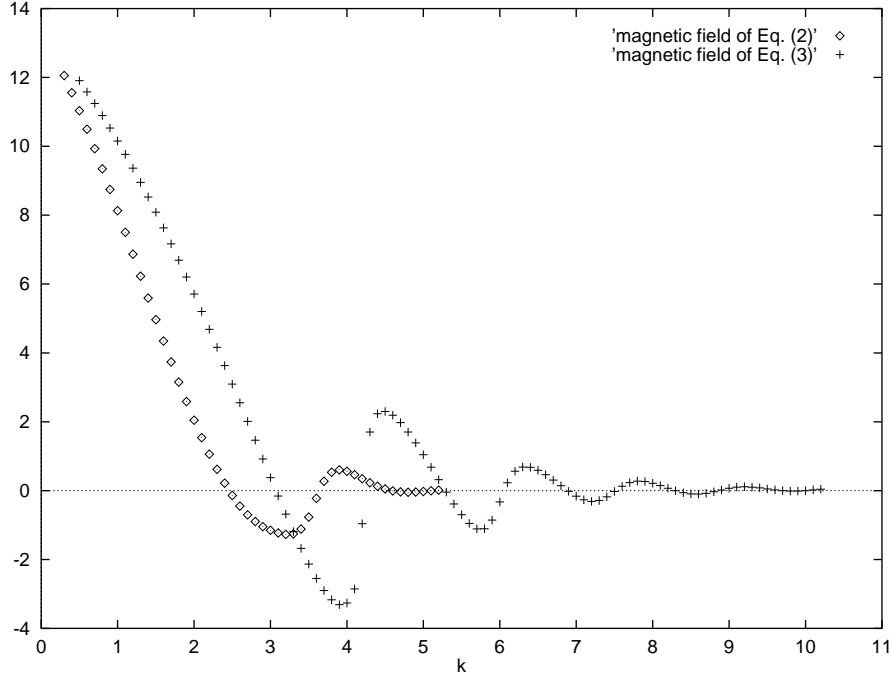,height=9cm,angle=-90}}}
 \caption { $\Delta(k)-c$ as a function of $k$ for
$\phi=4.5$
 and $d=1$} \label{1}
\end{figure}

In appendix B we describe two ways for the numerical calculation of the phase shifts
by solving an ordinary differential equation. The numerical calculation shows that
$\Delta(k)$ for large $k$ tends to a constant value $c=-\pi \phi^{2}$. An analytical
proof for this result can be obtained using the WKB approximation (22) for the phase
shift. In Fig. \ref{1} we have plotted the function $\Delta(k)-c$ for $\phi=4.5$ and
$d=1$. The numerical computation was performed for the magnetic fields $B_{1}$ and
$B_{2}$ of Eqs. (2) and (3) respectively. In order to achieve convergence of (20) for
$\phi=4.5$ we summed up to $l_{max}=20$ for the magnetic field $ B_{2}$ and to
$l_{max}=30$ for the Gaussian magnetic field $B_{1}$.

Fig. \ref{1} shows that the function $\Delta(k)-c$ tends rapidly to zero, apparently
like $1/k^{4}$ for the magnetic field $B_{2}$ and faster for the magnetic field
$B_{1}$, so the integral (19) is highly convergent. This is remarkable because it
means that the same function $\Delta(k)$ can be used for the $3+1$ dimensional case,
as can be seen from the equation (26).

The result that the function $\Delta(k)-c$ which corresponds to the magnetic field
$B_{1}$ tends faster to zero than the magnetic field $B_{2}$ suggests that in general
the function $\Delta(k)-c$ spreads when the corresponding magnetic field is getting
more localized. This has been checked by calculations for magnetic fields other than
$B_{1}$ and $B_{2}$ (see section 1). In addition we observed that when the magnetic
field is getting more localized we need less values of $l$ in order to achieve
convergence of (20).

The value of the function $\Delta(k)$ at $k=0$ is independent on the magnetic field
configuration and depends only on $\phi$. This becomes evident if we take into account
the Levinson theorem which is presented in appendix A. Thus the two curves in Fig.
\ref{1} intersect at $k=0$.

 Now from the relations (19) and (20) we obtain
\begin{eqnarray}
  E_{eff(2+1)}=-\frac{1}{2\pi}\int_{0}^{+\infty}\left(\sqrt{k^{2}+m^{2}_{f}}-m_{f}\right) \frac{d(\Delta(k)-c)}{dk}\:
dk
\end{eqnarray}
Integrating by parts we find
\begin{eqnarray}
  E_{eff(2+1)}=\frac{1}{2\pi}\int_{0}^{+\infty}\frac{k}{\sqrt{k^{2}+m^{2}_{f}}} (\Delta(k)-c) dk
\end{eqnarray}

For the $3+1$ dimensional case, from the relations (14),(18) and (20), again dropping
the free part from the density of states, we find
\begin{eqnarray}
E_{eff(3+1)}&=&\frac{1}{4\pi^{2}}\int_{0}^{+\infty}(k^{2}+m_{f}^{2})\ln\left(\frac{k^{2}+m_{f}^{2}}{m_{f}^{2}}\right)\frac{d(\Delta(k)-c)}{dk}
dk \nonumber \\ &
&-\frac{1}{4\pi^{2}}\int_{0}^{+\infty}k^{2}\frac{d(\Delta(k)-c)}{dk}dk
\end{eqnarray}
Integrating by parts we obtain
\begin{equation}
 E_{eff(3+1)}=-\frac{1}{2\pi^{2}}\int_{0}^{+\infty}k\ln\left(\frac{k^{2}+m^{2}_{f}}{m_{f}^{2}}\right) (\Delta(k)-c) dk
\end{equation}

\section{Dependence on the range $d$ for fixed magnetic flux $\phi$}

\begin{figure}[h]
\centerline{\hbox{\psfig{figure=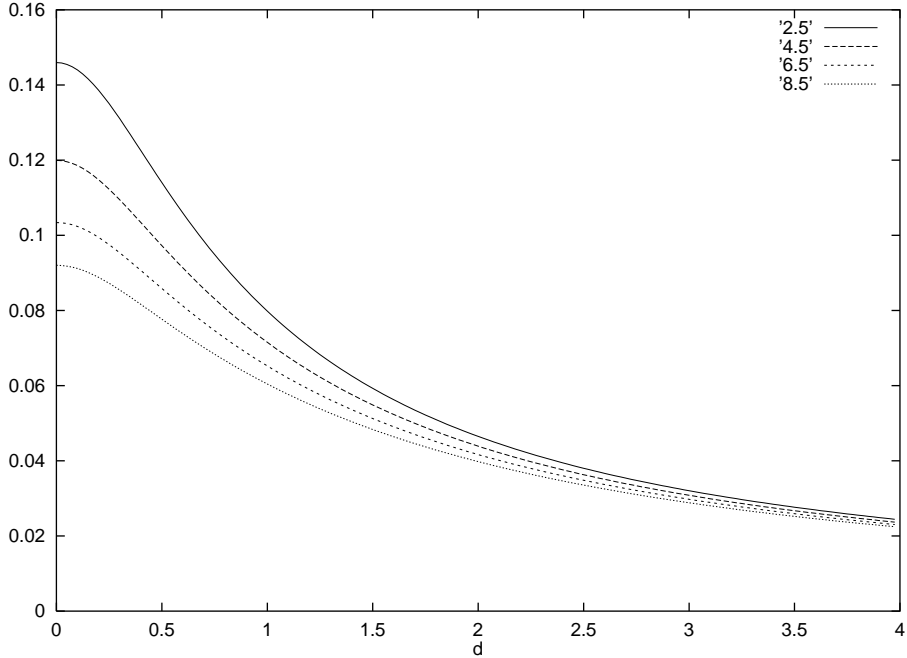,height=9cm,angle=-90}}} \caption
{$\phi^{-2}dE_{eff(2+1)}$ as a function of $d$ for $\phi=2.5,4.5,6.5,8.5$ and
$m_{f}=1$ for the Gaussian magnetic field of Eq. (2).} \label{2}
\end{figure}

\begin{figure}[h]
\centerline{\hbox{\psfig{figure=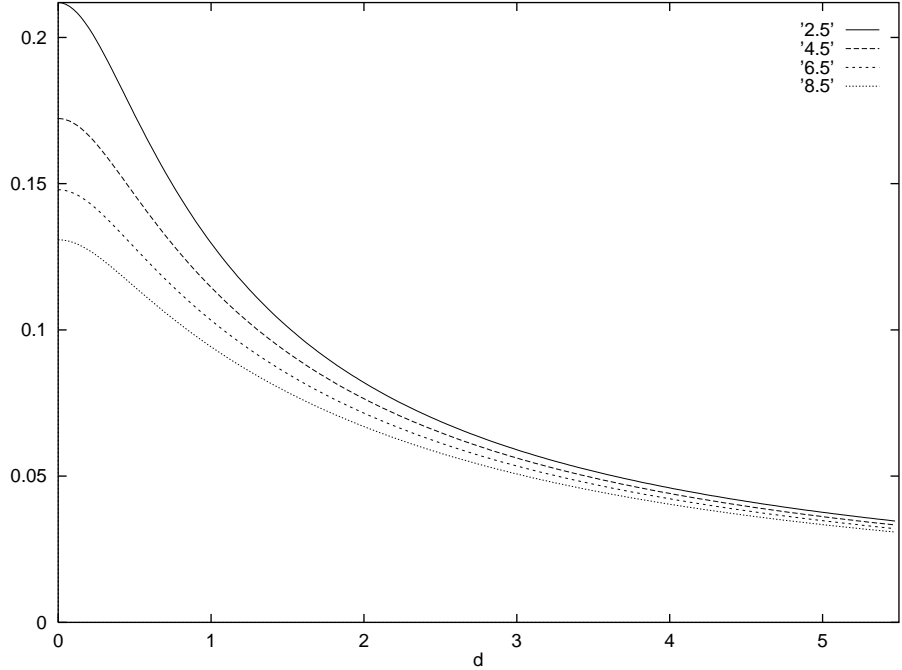,height=9cm,angle=-90}}} \caption {
$\phi^{-2}dE_{eff(2+1)}$ as a function of $d$ for $\phi=2.5,4.5,6.5,8.5$ and $m_{f}=1$
for the magnetic field of Eq. (3). } \label{3}
\end{figure}

A dimensional analysis shows that $\Delta(k)$ is a function of the form $F(k d,\phi)$.
If we make the change of variable $y=k d$ and set $\Delta(k)=F(k d,\phi)$, we find
\begin{eqnarray}
{E}_{eff(2+1)}&=&\frac{1}{2\pi
d}\int_{0}^{+\infty}\frac{y}{\sqrt{y^{2}+m_{f}^{2}d^{2}}}\;\left(F(y,\phi)-c\right)
dy
\\
 E_{eff(3+1)}&=&-\frac{1}{2\pi^{2}d^{2}}\int_{0}^{+\infty}y\ln\left(\frac{y^{2}+m_{f}^{2} d^{2}}{m_{f}^{2}d^{2}}\right) \left(F(y,\phi)-c\right)
 dy
\end{eqnarray}
where $d$ is a typical range of the magnetic field (see Eqs. (2)
and (3)).

In Fig. \ref{2} we see that $\eff$, for the Gaussian magnetic field $B_{1}$, is a
positive decreasing function of $d$ for fixed $\phi$. The asymptotic behavior of
$\eff$ for small values of $d$ is proportional to $1/d$. This can be seen from
relation (27). For large values of $d$ we are in the weak field regime and $\eff$ is
given approximately from the vacuum polarization diagram
\begin{eqnarray}
E_{eff(2+1)}^{(2)}=\frac{1}{8\pi^{2}d}\int_{0}^{1} dx x (1-x)
\int_{0}^{+\infty}dy
\frac{y|\widetilde{B}(y/d)|^{2}}{\sqrt{(m_{f}d)^{2}+x(1-x)y^{2}}}
\end{eqnarray}
where we have made the change of variable $q=y/d$ and $\widetilde{B}(q)$ is the
Fourier transform of the magnetic field. From the relation (29) we find that $\eff$ is
proportional to $1/d^{2}$ for large values of $d$.

The curves in Fig. \ref{2}  appear to approach each other as $d$ tends to infinity.
Thus the effective energy for large $d$ is proportional to $\phi^{2}$. This is
expected because the effective energy for large $d$ is given approximately by the
vacuum polarization diagram which is proportional to $\phi^{2}$.

\begin{figure}[h]
\centerline{\hbox{\psfig{figure=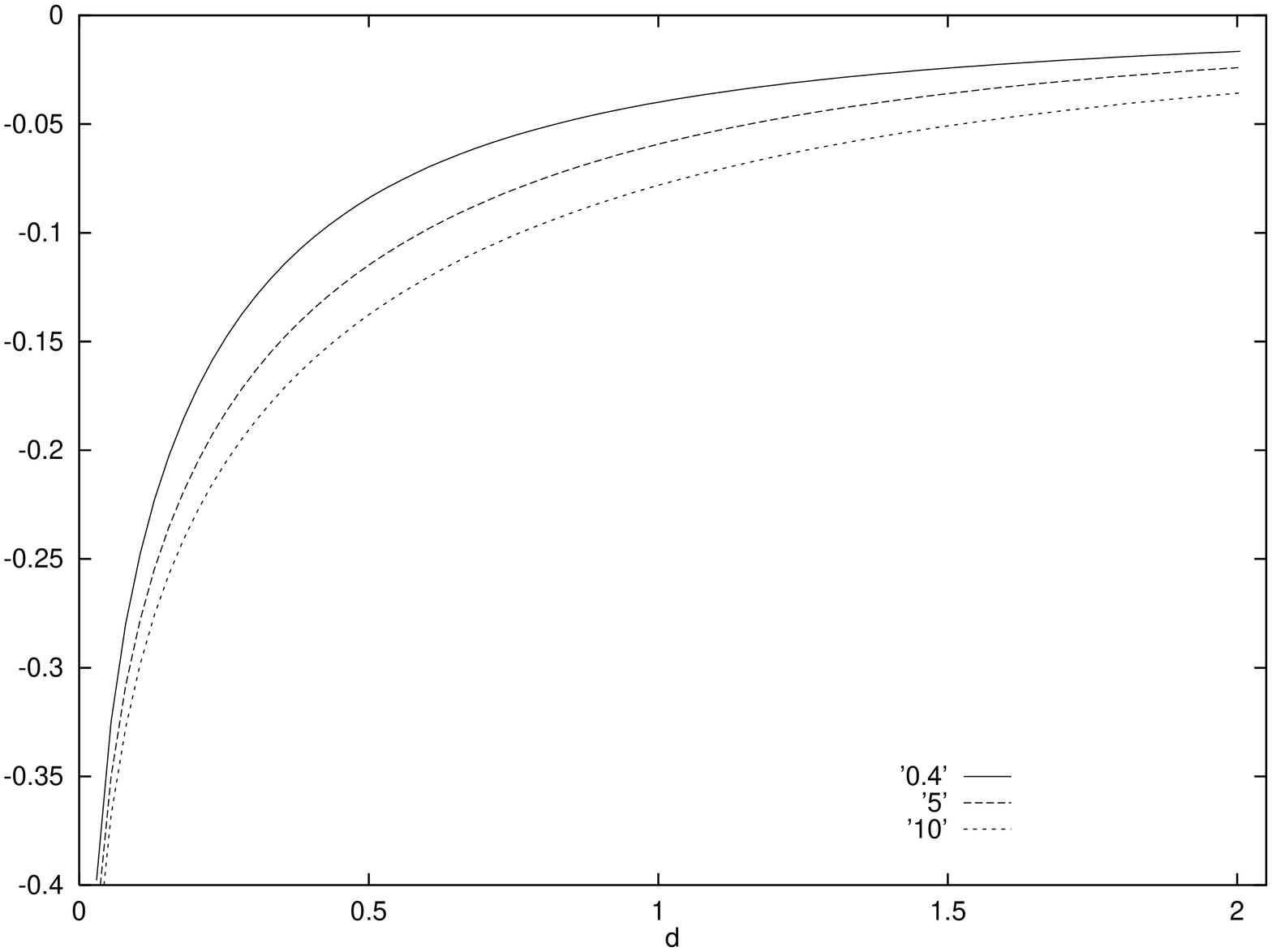,height=9cm,angle=-90}}} \caption {
$\phi^{-2}d^{2}E_{eff(3+1)}$ as a function of $d$ for $\phi=0.4,5,10$ and $m_{f}=1$
for the magnetic field of Eq. (3).} \label{4}
\end{figure}

In Fig. \ref{3} we present our numerical results for the magnetic field $B_{2}$. We
see that the effective energy as a function of $d$ for fixed $ \phi$ has the same
features as those of the Gaussian magnetic field of Eq. (2) \footnote{We have checked
that this is also true for the $3+1$ dimensional case. So the effective energy, in
$3+1$ dimensions, is a negative increasing function of $d$ for fixed magnetic flux (as
shown in Fig. \ref{4}) for all the magnetic fields of the form of the magnetic field
of Eq. (1).}. Also we can see that the effective energy of $B_{2}$ is always bigger
than the effective energy of $B_{1}$ for the same $\phi$ and $d$ (this is also true
for the corresponding classical energies). This fact seems to be a consequence of the
flux of the magnetic field $B_{2}$ being more concentrated than the flux of $B_{1}$.

The $3+1$ dimensional case, for the case of $B_{2}$, has already been studied with a
different method by M. Bordag and K. Kirsten in reference \cite{10}. Our results are
shown in Fig. \ref{4}. They agree closely with those of Fig. $3$ of reference
\cite{10}.

The $3+1$ dimensional effective energy, for large $d$, is given by
\begin{eqnarray}
E_{eff(3+1)}^{(2)}=-\frac{1}{8\pi^{3}d^{2}}\int_{0}^{1} dx x (1-x)
\int_{0}^{+\infty}dy y |\widetilde{B}(y/d)|^{2}
\ln(\frac{m_{f}^{2}d^{2}+y^{2}x (1-x)}{m_{f}^{2}d^{2}})
\end{eqnarray}
For $d\rightarrow+\infty$ we find
\begin{eqnarray}
E_{eff(3+1)}^{(2)}=-\frac{1}{240\pi^{3}m_{f}^{2}d^{4}}
\int_{0}^{+\infty}dy y^{3} |\widetilde{B}(y/d)|^{2}
\end{eqnarray}

For further discussion we will need the Fourier transforms of the magnetic fields of
Eqs. (2) and (3):
\begin{eqnarray}
 \widetilde{B}_{1}(q)&=&2 \pi \phi
\exp(-\frac{q^{2}d^{2}}{4})\\ \widetilde{B}_{2}(q)&=&4 \pi \phi
\frac{J_{1}(q d)}{q d}
\end{eqnarray}
From the relation (31) we see that for large $d$ the effective energy is proportional
to $1/d^{4}$. This is true only for magnetic fields for which the integral (31) over
$y$ is convergent. An example is the magnetic field $B_{1}$ with the Gaussian profile.
For the magnetic field $B_{2}$ the integral (31) is divergent. We found numerically
from (30) that in this case the asymptotic behavior is $O(1/d^{\alpha})$, where
$3<\alpha<4$.

For small values of $d$, from the relations (28) and (13) we find the following
asymptotic formula
\begin{eqnarray}
\effp=-\frac{1}{12\pi^{2}}\ln(\frac{1}{m_{f}d})\int d^{2}\vec{x}
B^{2}(\vec{x})
 \end{eqnarray}
The asymptotic formula (34) is the same as the one derived in reference \cite{10}.
From the relation (34) we obtain $\effp=\frac{-\phi^{2}}{3\pi
d^{2}}\ln(\frac{1}{m_{f}d})$ for the magnetic field $B_{2}$ and $
\effp=\frac{-\phi^{2}}{6\pi d^{2}}\ln(\frac{1}{m_{f}d})$ for the magnetic field
$B_{1}$.

The total energy for $B_{2}$ is given by
\begin{eqnarray}
E_{tot(3+1)}&=&E_{class(3+1)}+E_{eff(3+1)}\nonumber \\
            &=&\frac{2\pi\phi^{2}}{e^{2}d^{2}}+\frac{1}{d^{2}}f(m_{f}d,\phi)
\end{eqnarray}
where we have set $E_{eff(3+1)}=\frac{1}{d^{2}}f(m_{f}d,\phi)$.
The physical meaning of the total energy is the energy we should
spend in order to create the magnetic flux tube. In this article
we have computed this energy at one loop order (i.e we have not
taken into account virtual photons). This approximation is valid
only for small values of the coupling constant $e$.

For small values of $d$ we find
\begin{equation}
E_{tot(3+1)}=\frac{2\pi\phi^{2}}{e^{2}d^{2}}-\frac{\phi^{2}}{3\pi
d^{2}}\ln(\frac{1}{m_{f}d})
\end{equation}
We see that the logarithm, which comes from the effective energy part, dominates for
small values of $d$. This asymptotic behavior of the total energy is not reliable. The
reason is that small values of $d$ correspond to magnetic fields whose Fourier
transforms have contributions from high momenta at which the running coupling constant
can not be assumed small.

The asymptotic relation (36) for the total energy can be written in the form
\begin{eqnarray}
E_{tot(3+1)}=\frac{2\pi\phi^{2}}{e^{2}_{eff}(d)d^{2}}
\end{eqnarray}
where we have defined a d-dependent effective coupling constant,
for small values of $d$, according to
\begin{eqnarray}
\frac{1}{e^{2}_{eff}(d)}=\frac{1}{e^{2}}-\frac{1}{6\pi^{2}}\ln(\frac{1}{m_{f}d})
\end{eqnarray}
It is interesting to compare this to the B-dependent effective
coupling constant of reference \cite{eff1}, for a homogeneous
magnetic field, which for large $B$ reads
\begin{eqnarray}
\frac{1}{e^{2}_{eff}(B)}=\frac{1}{e^{2}}-\frac{1}{12\pi^{2}}\ln(\frac{B}{m_{f}^{2}})
\end{eqnarray}

\section{Dependence on the range $d$ for fixed magnetic field strength $B_{m}=2\phi/d^{2}$}

\begin{figure}[h]
\centerline{\hbox{\psfig{figure=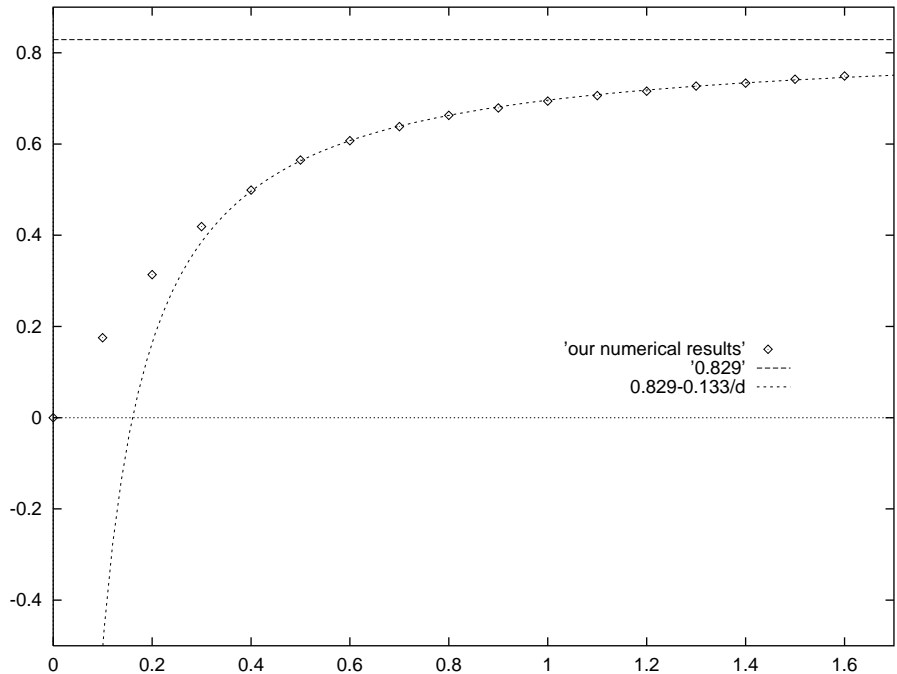,height=9cm,angle=-90}}}
\caption {$ E_{eff(2+1)}/d^{2}$ as a function of $d$ for $B_{m}=5$
and $m_{f}=1$ for the magnetic field of Eq. (3). The discrete
points correspond to our numerical results and the continuous line
to the best fit curve of the form $a_{0}-a_{1}/d$. The asymptotic
value $a_{0}=0.829\pm 0.002$ agrees quite well with that
calculated from the Schwinger formula: $0.829$. Our numerical
results for $a_{0}$ and $a_{1}$ are rounded to three decimal
points.} \label{5}
\end{figure}

\begin{figure}[h]
\centerline{\hbox{\psfig{figure=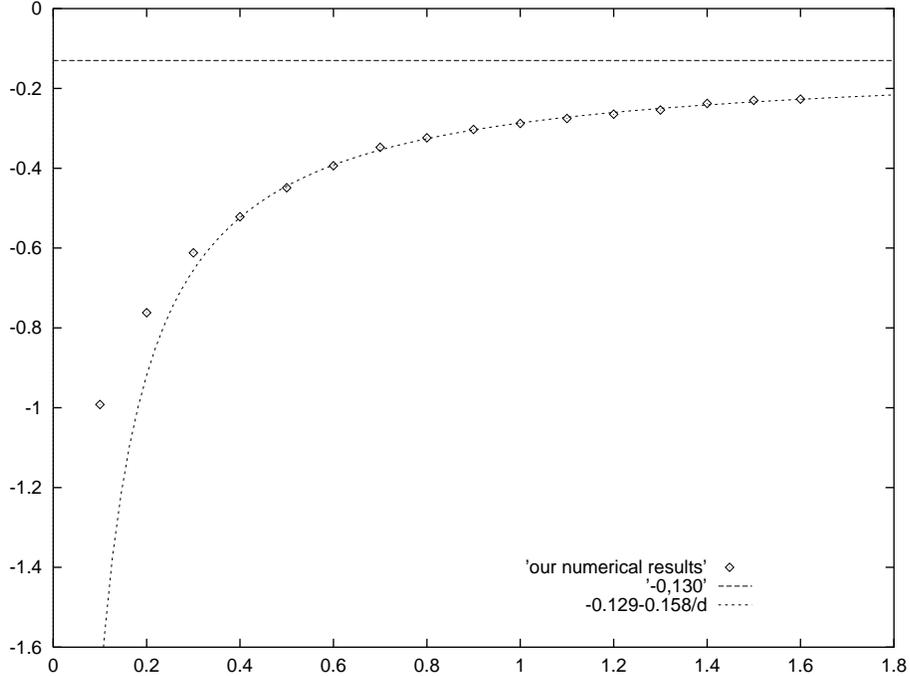,height=9cm,angle=-90}}}
\caption {$ E_{eff(3+1)}/d^{2}$ as a function of $d$ for $B_{m}=5$
and $m_{f}=1$ for the magnetic field of Eq. (3). The discrete
points correspond to our numerical results and the continuous line
to the best fit curve of the form $a_{0}-a_{1}/d$. The asymptotic
value $a_{0}=0.129\pm 0.002$ agrees quite well with that
calculated from the Schwinger formula: $0.130$. Our numerical
results for $a_{0}$ and $a_{1}$ are rounded to three decimal
points.} \label{6}
\end{figure}

It is convenient to define a characteristic magnetic field
strength as $B_{m}=\int\vec{B}\cdot d\vec{S}/\pi d^{2}$, in order
to estimate the intensity of the magnetic field, where $d$ is the
range of the magnetic field. In the previous section we examined
the dependence on $d$ of the effective energy for fixed magnetic
flux $\phi$. Here, we study the dependence on $d$ (or $m_{f}d$)
for fixed magnetic field strength $B_{m}=2\phi/d^{2}$ (or
$B_{m}/m_{f}^{2}$).

First, we examine the magnetic field of Eq. (3) (the Gaussian
magnetic field of Eq. (2) is examined in the next section). An
interesting feature of this field (of Eq. (3)), is that when its
range $d$ increases it tends to a homogeneous magnetic field.
Thus, in the homogeneous limit ($d\gg1/\sqrt{B_{m}}$), we expect
our numerical results to approach the result calculated from the
Schwinger formula , which for the cases of $2+1$ and $3+1$
dimensions reads
\begin{equation}
E_{eff(2+1)}^{Sch}={\cal A} \: \frac{B^{3/2}}{8
\pi^{3/2}}\int_{0}^{+\infty}\frac{ds
}{s^{3/2}}\left(\coth(s)-\frac{1}{s} \right) e^{-s m_{f}^{2}/B}
\end{equation}
\begin{eqnarray}
E_{eff(3+1)}^{Sch}= {\cal A} \:\frac{B^{2}}{8 \pi^{2}}
\int_{0}^{+\infty}
\frac{ds}{s^{2}}\left(\coth(s)-\frac{1}{s}-\frac{s}{3}\right)e^{-sm_{f}^{2}/B}
\end{eqnarray}
where ${\cal A}$ is the area of the space box on the $x-y$ plane and $B$ is the
magnetic field strength of the homogeneous magnetic field.

In Fig. \ref{5} we have plotted the effective energy divided by
$d^{2}$ for $B_{m}=5$ and $m_{f}=1$ for the magnetic field of Eq.
(3). We see that $E_{eff(2+1)}/d^{2}$ is a positive increasing
function of $d$ which tends $\mathit{very \; slowly}$ to an
asymptotic value. The determination of this asymptotic value is
not possible with a directly numerical computation, as for large
values of $d$ the numerical error of our method becomes
significant. However, we see in Fig. \ref{5} that our numerical
results (for $d\geq 0.4$) lie on a curve of the form
$a_{0}-a{1}/d$. Fitting a curve of this form to our data we obtain
$a_{0}=0.829\pm 0.002$ and $a_{1}=0.133\pm 0.001$. The asymptotic
value $a_{0}=0.829\pm 0.002$ agrees quite well with that
calculated from the Schwinger formula:
$E^{Sch}_{eff(2+1)}/d^{2}=0.829$. This asymptotic behavior has
been also observed for other values of $B_{m}$ and $m_{f}$ which
cover all the range of the characteristic ratio $B_{m}/m^{2}_{f}$,
that determines how strong is the magnetic field.

For small values of $d$, if we take into account the vacuum
polarization diagram of Eq. (29) which is good approximation for
the effective energy, we find that $E_{eff(2+1)}/d^{2}$ is
proportional to $d$. This is in agreement with the linear behavior
of $E_{eff(2+1)}/d^{2}$, for $d\rightarrow 0$, which is seen in
Figs. \ref{5} and \ref{7}.

In the case of $3+1$ dimensions, we see in Fig. \ref{6} that our
numerical results are negative and tend slowly to an asymptotic
value, which also agrees closely with that calculated from the
Schwinger formula of Eq. (41). The asymptotic behavior for large
$d$ is the same as that observed in the case of $2+1$ dimensions
($a_{0}-a_{1}/d$).

For small $d$ we obtain from the vacuum polarization diagram of
Eq. (30) that $E_{eff(2+1)}/d^{2}$ is proportional to $\ln d$. We
see that this logarithmic behavior is not clear in Fig. \ref{7}
since we have not plotted enough points for small values of $d$.
We avoided this, as in the region of small $d$ $(d\approx 0.1)$
the numerical accuracy we could achieve was not satisfactory.

A different way, of finding the behavior of the effective energy
for small $d$, is to replace $F(y,\phi)=\phi F^{(1)}(y)+\phi^{2}
F^{(2)}(y)$ in Eqs. (27) and (28) and take into account that the
linear term is identically zero ($F^{(1)}(y)=0$). We can prove
that $F^{(1)}(y)=0$ using the first Born approximation for the
phase shifts.

Our numerical work shows there is a significant possibility that
the $1/d$ term in the above-mentioned asymptotic behavior
($a_{0}-a_{1}/d$) is true. If, indeed, this behavior is true, it
would be interesting if one could gave an analytical proof.

Note, that this asymptotic behavior is expected only for
discontinuous magnetic fields like the one of Eq. (3). In Fig.
\ref{7} we see that in the case of the Gaussian magnetic field of
Eq. (2), our results for large $d$ have the same asymptotic
behavior with that of the derivative expansion:
$b_{0}-b_{1}/d^{2}$. This means that there is a remarkable
difference between smooth and discontinuous magnetic fields, in
the way their effective energies (divided by $d^{2}$) tend to
their asymptotic values. It is obvious that this difference, is
related to the fact that the derivative expansion fails for
discontinuous magnetic fields, like the magnetic field of Eq. (3).

\section{Derivative expansion}

\begin{figure}[h]
\centerline{\hbox{\psfig{figure=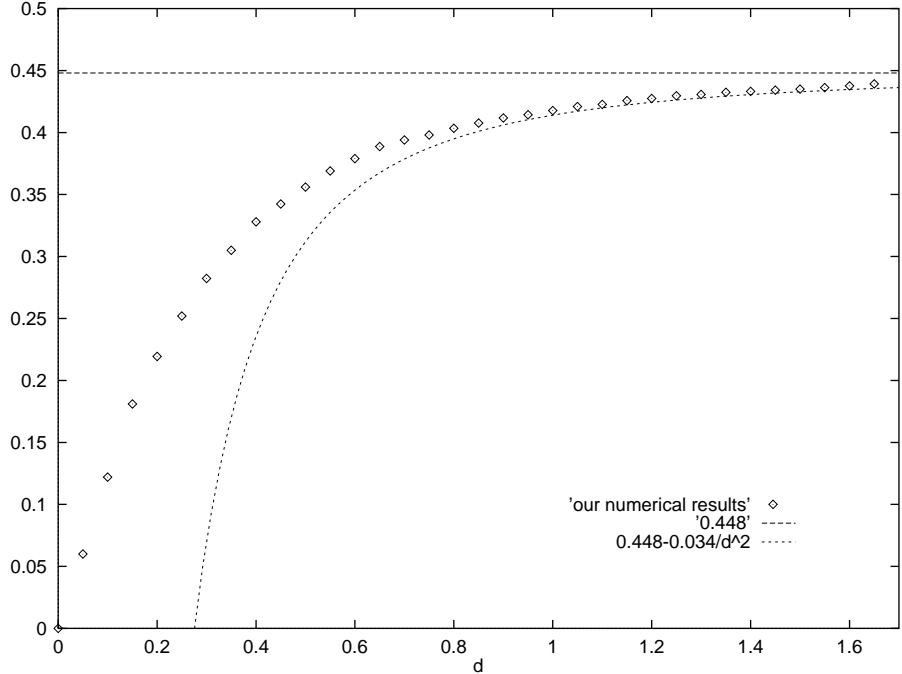,height=9cm,angle=-90}}} \caption {$
E_{eff(2+1)}/d^{2}$ as a function of $d$ for $B_{m}=5$ and $m_{f}=1$ for the Gaussian
magnetic field of Eq. (2). The discrete points correspond to our numerical results and
the continuous line $0.448-0.034/d^{2}$ to derivative expansion.}\label{7}
\end{figure}

An approximate way to compute the effective action in the presence
of a static magnetic field, is the derivative expansion. This
method should give accurate results for slowly varying magnetic
fields. Our aim is to examine cases in which the derivative
expansion fails to provide a good approximation. Also, we can test
our numerical results comparing with those of derivative expansion
in the homogeneous limit ($B_{m}d^{2}\gg1$), where the derivative
expansion is expected to give accurate results.

The derivative expansion for $3+1$ and $2+1$ dimensional QED has
been investigated in references \cite{der3,der2} respectively. For
the case of $2+1$ dimensions, if we keep the first two terms (a
zeroth order term plus a first order correction) the effective
energy for a static magnetic field is given by
\begin{eqnarray}
 E_{eff(2+1)}^{der}=E_{(2+1)}^{(0)}+ E_{(2+1)}^{(1)}
\end{eqnarray}
where
\begin{eqnarray}
 E_{(2+1)}^{(0)}=\int d^{2}\vec{x}\frac{B^{\frac{3}{2}}}{8\pi^{\frac{3}{2}}}\int_{0}^{+\infty}\frac{1}{s^{\frac{3}{2}}}(\coth(s)-\frac{1}{s}) \:
 e^{-s \: m_{f}^{2}/B} ds
\end{eqnarray}
\begin{eqnarray}
E_{(2+1)}^{(1)} =\frac{1}{8}(\frac{1}{4\pi})^{\frac{3}{2}}\int
d^{2}\vec{x} \frac{(\nabla
B)^{2}}{B^{\frac{3}{2}}}\int_{0}^{+\infty}\frac{1}{s^{\frac{1}{2}}}
e^{-s \: m_{f}^{2}/B} (\frac{d}{ds})^{3}[s\coth(s)] ds
\end{eqnarray}
For the special case of static magnetic fields, the above formulas
were obtained in references \cite{der4,der5}.

In Fig. \ref{7} we have plotted the effective energy
$E_{eff(2+1)}/d^{2}$ as a function of $d$, for fixed magnetic
field strength $B_{m}=5$ and $m_{f}=1$ for the Gaussian magnetic
field $B_{1}$. We see, as expected, that in the homogeneous limit
$(d\gg 1/\sqrt{B_{m}})$, our results agree quite well with those
of the derivative expansion. In the small $d$ limit $(d \ll
1/\sqrt{B_{m}})$, the derivative expansion fails to approximate
the effective energy. See the deviations in Fig. \ref{7} for $d
<0.8$.

The case of $3+1$ dimensions has not been included, as it is similar with the case of
$2+1$ dimensions.

\section{The strong magnetic field ($B_{m}\gg m_{f}^{2}$)}

In order to study the strong magnetic field ($B_{m}\gg m_{f}^{2}$)
behavior of the effective energy, we set
$\Delta(k)=G(kB_{m}^{-\frac{1}{2}},B_{m} d^{2})$ and make the
change of variable $y=k B_{m}^{-1/2}$ in the relations (26) and
(24)
\begin{eqnarray}
  E_{eff(2+1)}=\frac{B_{m}^{\frac{1}{2}}}{2\pi}\int_{0}^{+\infty}\frac{y}{\sqrt{y^{2}+m^{2}_{f}/B_{m}}} \left(G(y,B_{m} d^{2})-c\right)
  dy
\end{eqnarray}

\begin{equation}
 E_{eff(3+1)}=-\frac{B_{m}}{2\pi^{2}}\int_{0}^{+\infty}y\ln\left(\frac{y^{2}+m^{2}_{f}/B_{m}}{m^{2}_{f}/B_{m}}\right)\left(G(y,B_{m} d^{2})-c\right)
  dy
\end{equation}
where $B_{m}$ is a characteristic magnetic field strength defined
as $B_{m}=\Phi/\pi d^{2}$ and $\Phi$ is the magnetic flux.

 For $B_{m}\gg m^{2}_{f}$ we find the asymptotic expressions
\begin{eqnarray}
  E_{eff(2+1)}=\frac{B_{m}^{\frac{1}{2}}}{2\pi}\int_{0}^{+\infty} \left(G(y,B_{m} d^{2})-c\right)
  dy
\end{eqnarray}

\begin{equation}
 E_{eff(3+1)}=-\frac{B_{m}}{2\pi^{2}}\ln\left(B_{m}/{m^{2}_{f}}\right)\int_{0}^{+\infty}y\left(G(y,B_{m} d^{2})-c\right)
  dy
\end{equation}

In the case of $3+1$ dimensions, if we take into account the
equation (13), we obtain an asymptotic formula for the case of
strong magnetic field.
\begin{equation}
 E_{eff(3+1)}=-\frac{1}{24\pi^{2}}\ln\left(B_{m}/{m^{2}_{f}}\right)\int_{0}^{+\infty}d^{2}\vec{x}B^{2}(\vec{x})
\end{equation}
The above equation is more general than Eq. (34), which was
obtained in the limit $m_{f}d\rightarrow 0$ for fixed $\phi$.

For fixed $m_{f}d$, Eq. (49) yields an asymptotic formula in the
case of large magnetic flux $\phi$
 \begin{equation}
 E_{eff(3+1)}=-\frac{1}{24\pi^{2}}\ln\left(2\phi\right)\int_{0}^{+\infty}d^{2}\vec{x}B^{2}(\vec{x})
\end{equation}

From Eq. (50) we obtain $E_{eff(3+1)}=\frac{-1}{6\pi
d^{2}}\phi^{2}\ln(2\phi)$ for the magnetic field of Eq. (3) and
$E_{eff(3+1)}=\frac{-1}{12\pi d^{2}}\phi^{2}\ln(2\phi)$ for the
magnetic field of Eq. (2). We see that for large $\phi$ the $3+1$
dimensional effective energy is proportional to $-
\phi^{2}\ln\phi$. Also, this analytical result explains the weak
dependence on $\phi$ of the $E_{eff(3+1)}/\phi^{2}$ for the
magnetic field of Eq. (3), that was observed numerically in
reference \cite{10}.

\begin{figure}[h]
\centerline{\hbox{\psfig{figure=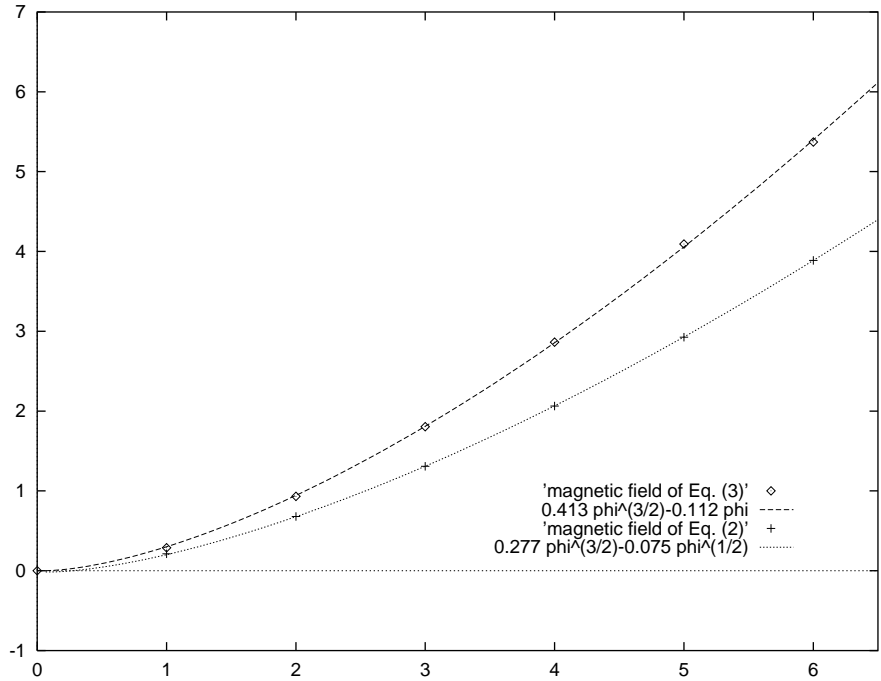,height=9cm,angle=-90}}}
\caption {$E_{eff(2+1)}$ as a function of $\phi$ for $d=1$ and
$m_{f}=0$ for the magnetic fields of Eqs. (2) and (3). The
discrete points correspond to our numerical results and the
continuous lines to the best fit curves of the form
$w_{0}\phi^{3/2} -w_{1}\phi$ for the magnetic field of Eq. (3) and
$v_{0}\phi^{3/2} -v_{1}\phi^{1/2}$ for the magnetic field of Eq.
(2). Our numerical results for the coefficients are rounded to
three decimal points.} \label{8}
\end{figure}

In the case of $2+1$ dimensions, the asymptotic formula for $\phi\rightarrow +\infty$
is obtained from Eq. (45) by setting $m_{f}=0$. In Fig. \ref{8} we have plotted the
effective energy as a function of $\phi$ for $d=1$ and $m_{f}=0$, for the magnetic
fields of Eqs. (2) and (3). Also, in the same figure we see that the curves
$w_{0}\phi^{3/2}-w_{1}\phi$ and $v_{0}\phi^{3/2} -v_{1}\phi^{1/2}$, which correspond
to the magnetic fields of Eq. (2) and (3), fit our data very well. Thus, for large
$\phi$ the $2+1$ dimensional effective energy is proportional to $\phi^{3/2}$. Also,
we observe that the coefficients $w_{0}=0.413 \pm 0.008$ and $v_{0}=0.2769 \pm 0.0005$
agree closely with the coefficients obtained from the formula
\begin{equation}
E_{eff(2+1)}=-\frac{\zeta(-1/2)}{\sqrt{2}\pi}\int d^{2}\vec{x}\:
B^{3/2}(\vec{x})
\end{equation}
which are $0.416$ for the magnetic field of Eq. (3) and $0.2772$
for the magnetic field of Eq. (2). This formula is obtained from
the leading term of the derivative expansion (given by the Eq.
(43)) if we set $m_{f}=0$. Our numerical results show that the
formula (51) gives the asymptotic behavior for large $\phi$, in
the case of $2+1$ dimensions. Note that the formula (51) does not
include all the cases of strong magnetic field $B_{m}=2
\phi/d^{2}$. In the case of small $m_{f}d$ and fixed $\phi$ it is
not correct. In this case the effective energy is proportional to
$1/d$ as shown by Eq. (27) in section 4.

\section{Conclusions}

We presented a numerical study of the fermion-induced $3+1$ and
$2+1$ QED effective action in the presence of any static
cylindrically symmetric magnetic field with finite magnetic flux.
We used a simple novel approach for the fast computation of the
effective action. The numerical work was greatly facilitated by
the observation that the function $\Delta(k)-c$ (see Eq. (20))
tends rapidly to zero for large $k$. The integrals (27) and (28)
for the effective energy for $2+1$ and $3+1$ dimensional QED
respectively are rendered convergent due to this fact.

The function $\Delta(k)-c$ is the only quantity we need to know in order to compute
the effective energy (see Eqs. (27) and (28)). We have performed computations for
eight magnetic field configurations (see section 1). The form \footnote{see section 3
and Fig. \ref{1}} of the function $\Delta(k)-c$ is similar for all these fields. This
means that our conclusions for the effective energy are quite general and are not
valid just for the magnetic fields of Eqs. (2) and (3).

For $e^{2}/4\pi=1/137$, our numerical work shows that for the families of magnetic
fields we examined, there is no radius $d$ of the magnetic flux tube that minimizes
the total energy for fixed magnetic flux. The main reason for this is the small
contribution (of the order of one per cent) of the effective energy to the total
energy. An exception, where the contribution is not small, is the case of $3+1$
dimensional effective energy for small $d$, as may be seen from Eq. (36).

In section 5, we observed that our numerical results for $E_{eff}/d^{2}$, in the case
of the magnetic field of Eq. (3), converge slowly to an asymptotic value given by the
Schwinger formula for a homogeneous magnetic field. In addition, we showed numerically
that this asymptotic behavior for large $d$ is of the form $a_{0}-a_{1}/d$. The case
of the Gaussian magnetic field of Eq. (2) is different, as the convergence to the
asymptotic value is quite faster than the case of the magnetic field of Eq. (3). The
asymptotic behavior of $E_{eff}/d^{2} \sim b_{0}-b_{1}/d^{2}$ in this case, as we see
in Fig. \ref{7}, is the same with that given by the derivative expansion.
Generalizing, we expect that this behavior would be valid for all smooth magnetic
fields of the form we examine. In the case of discontinuous magnetic fields, for which
the derivative expansion fails, we expect a different asymptotic behavior of the form:
$a_{0}-a_{1}/d$.

We studied the case of large magnetic flux $\phi$ (strong magnetic field) when the
spatial size of the magnetic flux tube is fixed. We showed that the effective energy
for large $\phi$ is proportional to $-\phi^{2} \ln\phi$ in the case of $3+1$
dimensions and to $\phi^{3/2}$ in the case of $2+1$ dimensions. Especially in the case
of $3+1$ dimensions, we derived an explicit formula for large magnetic flux $(\phi\gg
1)$.

It would be interesting, if for large $\phi$ the effective energy in the presence of
an inhomogeneous magnetic field would become greater than the classical energy. In
view of the above-mentioned results we should exclude this possibility at least for
the class of inhomomogeneous magnetic fields of the form of a flux tube. An exception
may be the case of $3+1$ dimensions for very large $\phi$. However, for very large
$\phi$ the asymptotic behavior $-\phi^{2} \ln\phi$ is not reliable, as the
contributions of the higher loop corrections (two-loop corrections, three-loop
corrections etc) to the QED effective energy cannot be assumed small.

\section{Acknowledgements}
I am grateful to Professor G. Tiktopoulos, who proposed this topic to me and
supervised the work. I also would like to thank him for numerous valuable discussions
and the careful reading of the manuscript, as well as Dr. J. Alexandre, and Professors
K. Farakos, G. Koutsoumbas and A. Kehagias for useful comments and suggestions.
Partially supported by NTUA program Archimides.

\appendix

\section{APPENDIX: Effective energy for massless fermions}

In this appendix we derive an expression for the effective energy
for massless fermions, since Eq. (14) contains a singularity at
$m_{f}=0$. It is remarkable, that this singularity does not exist
in the unrenormalized effective energy of Eq. (12). It arises due
to the renormalization procedure used to remove the ultraviolet
divergent part of the effective energy. However, the
above-mentioned singularity does not appear, if instead of the
renormalization condition $\Pi(0)=0$ we use the off-shell
renormalization condition $\Pi(-M^{2})=0$.

Setting $m_{f}=0$ in the unrenormalized effective energy of Eq. (12) we find
\begin{eqnarray}
E_{eff(3+1)}&=&\frac{L_{z}}{4\pi}\sum_{\{n\}}E_{\{n\}}\ln(\frac{E_{\{n\}}}{\Lambda^{2}})\nonumber
-\frac{1}{4\pi}\sum_{\{n\}}E_{\{n\}}
\end{eqnarray}

The ultraviolet divergent part in the above equation is removed if we include the
following counterterm in the QED Lagrangian.
\begin{equation}
S_{count}=(Z_{3}-1)\int d^{4}x(-\frac{1}{4}F_{\mu\nu}F^{\mu\nu})
\end{equation}
where
\begin{equation}
Z_{3}=1-\frac{1}{2 \pi^{2}}\left(\frac{1}{6}
\ln(\frac{\Lambda^{2}}{M^{2}})+\frac{5}{18}\right)
\end{equation}
This counterterm is defined uniquely by the off-shell renormalization condition
$\Pi(-M^{2})=0$. In this case the renormalized effective action for massless fermions
is
\begin{eqnarray}
E_{eff(3+1)}^{(ren)}=\frac{L_{z}}{4\pi}\sum_{\{n\}}E_{\{n\}}\ln(\frac{E_{\{n\}}}{M^{2}})+\frac{L_{z}}{6\pi}\sum_{\{n\}}E_{\{n\}}
\end{eqnarray}

In a similar way with that of section $3$, we obtain
\begin{eqnarray}
  E_{eff(3+1)}^{(ren)}=\frac{1}{d^{2}}\left(c_{1}(\phi)\ln(d M)+c_{2}(\phi)\right)
\end{eqnarray}
where
\begin{eqnarray}
&&c_{1}(\phi)=\frac{1}{\pi^{2}}\int_{0}^{+\infty}y(F(y,\phi)-c)dy=\frac{d^{2}}{12\pi^{2}}\int
\vec{B}^{2}(\vec{x})d^{2}\vec{x}
\\
&&c_{2}(\phi)=-\frac{1}{\pi^{2}}\int_{0}^{+\infty}y\ln(y)(F(y,\phi)-c)dy-\frac{5}{6}c_{1}(\phi)
\end{eqnarray}

The total energy is given by the equation
\begin{eqnarray}
E_{tot}&=&E_{class}+E_{eff}\nonumber \\
&=&\frac{1}{2e^{2}_{M}}\int
\vec{B}^{2}(\vec{x})d^{2}\vec{x}+\frac{1}{d^{2}}\left(c_{1}(\phi)\ln(d
M)+c_{2}(\phi)\right)
\end{eqnarray}
where  $e_{M}$ is the charge of the fermion defined at the scale $M$. The
transformation law for the charge $e_{M}$ is given by the equation
\begin{equation}
\frac{1}{e_{M'}^{2}}=\frac{1}{e_{M}^{2}}-\frac{1}{6\pi^{2}}\ln\left(\frac{M'}{M}\right)
\end{equation}
which also guarantees that the total energy is independent from
the renormalization scale $M$.

\section{APPENDIX: The appearance of metastable states for large magnetic flux $\phi$}

\begin{figure}[h]
\centerline{\hbox{\psfig{figure=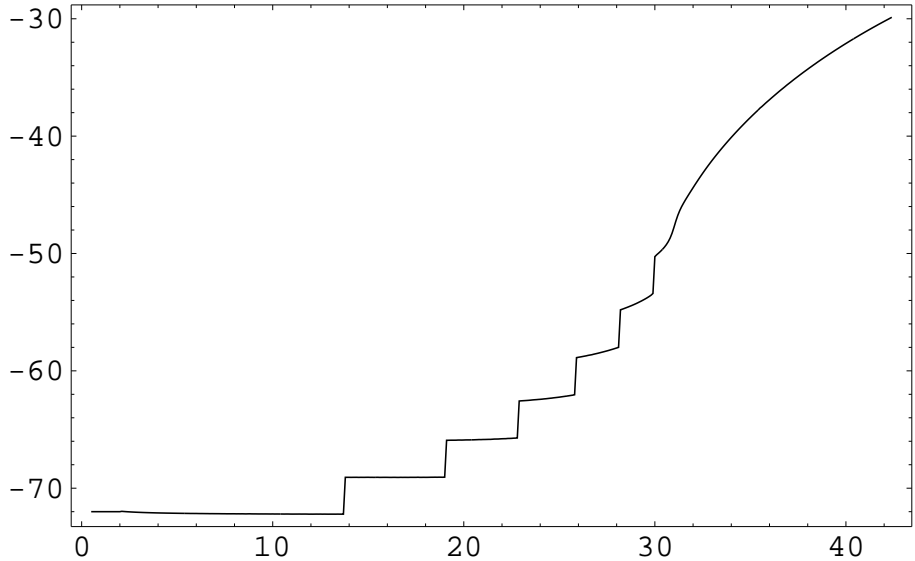,height=7.5cm,angle=0}}} \caption {
$\delta_{l,\sigma}(k)$ as a function of $k$ for $l=1$, $\sigma=1$, $d=1$ and $\phi=50$
for the Gaussian magnetic field of Eq. (2)} \label{9}
\end{figure}

\begin{figure}[h]
\centerline{\hbox{\psfig{figure=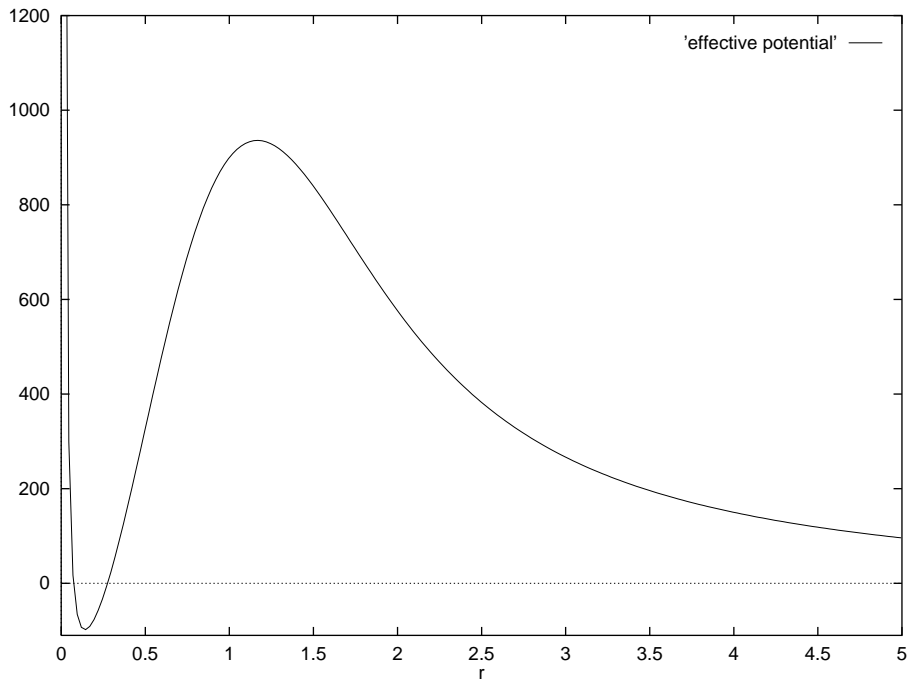,height=9cm,angle=-90}}} \caption {
$v_{l,\sigma}(r)+\frac{l^{2}-1/4}{r^{2}}$ as a function of $r$ for $l=1$, $\sigma=1$,
$d=1$ and $\phi=50$ for the Gaussian magnetic field of Eq. (2)} \label{10}
\end{figure}

In this appendix we examine the appearance of metastable states for large magnetic
flux $\phi$. Note that in the large magnetic flux $\phi=B_{m}d^{2}/2$ limit we
distinguish two cases: a) a large magnetic field strength $B_{m}$ keeping $d$ constant
and b) a long range $d$ of the magnetic field keeping $B_{m}$ constant.

In Fig. \ref{9}, we see that the function $\delta_{l,s}(k)$ exhibits jumps at a finite
number of values of $k$. Every jump increases the phase shift by $\pi$ and corresponds
to a metastable state of the electron. It is remarkable that these jumps occur in a
small but nonzero interval $\Delta k$ at which the density of states appears as a very
sharp peak. For the calculation of the phase shifts in this case we have used the
differential equation (75) of appendix B, which is appropriate for large $\phi$. For
the plot we calculated the phase shifts with a step equal to $0.1$. It may be noted,
incidentally that our numerical results for the phase shifts at $k=0$ are in agreement
with the Levinson's theorem \cite{17}. Levinson's theorem, for a two dimensional
Schrodinger equation and for a potential with asymptotic behavior $O(1/r^{2})$ for
large $r$, has been investigated in reference \cite{19}. Thus for our special case
\begin{eqnarray} \delta_{l,s}(0)=
    \left\{\begin{array}{cl}
      \pi+\frac{\pi}{2}(|l|-|l-\phi|) &
      \mbox{if a zero mode exists}
        \\ \frac{\pi}{2}(|l|-|l-\phi|) & \mbox{otherwise}
  \end{array}\right.
\end{eqnarray}

The metastable states are due to the potential well form of the effective potential
$v_{l,\sigma}(r)+(l^{2}-1/4)/r^{2}$ as shown in Fig. \ref{10}. It is interesting to
compare with the WKB approximation to the eigenvalues $k_{n}^{2}$ of the metastable
states as roots of the equation
\begin{equation}
\int_{a}^{b}\left(\sqrt{k_{n}^{2}-v_{l,s}(r)-\frac{l^{2}}{r^{2}}}\right)
dr=(n+\frac{1}{2})\pi \quad n=0,1,2...
\end{equation}
where $a$ and $b$ are the turning points. In Table \ref{a} we see that our numerical
results agree closely with those of WKB approximation. Note, that the electron
metastable state energies corresponding to the eigenvalues $k_{n}^{2}$, are
$\sqrt{k_{n}^{2}+m_{f}^{2}}$.

\begin{table}[h]
\begin{center}
 \begin{tabular}{|c|c|c|} \hline
 $n$ &WKB approximation $k_{n}$  &Numerical computation \\ \hline
 1   & 13.721 & 13.7-13.8 \\
 2  &19.051  &19.0-19.1  \\
 3   &22.881   &22.8-22.9   \\
 4   &25.854   &25.8-25.9   \\
 5   &28.200   &28.1-28.2\\
 6   &29.997 & 29.9-30.0\\  \hline
   \end{tabular}
\end{center}
 \caption{We compare the energies of the metastable states as they are obtained from the
  Fig. \ref{8} with the WKB approximation results for $d=1$, $l=1$, $\sigma=1$ and $\phi=50$.}
   \label{a}
\end{table}

\section{APPENDIX: The role of the Landau levels for the effective action}

Taking into account the Eqs. (9) and (14), it is obvious, that the effective energy is
determined uniquely by the spectrum of the eigenvalue equation
$(\gamma^{m}D_{m})^{2}\Psi_{\{n\}}(x)=E_{\{n\}}\Psi_{\{n\}}(x)$. In the section 5, it
was shown numerically, that the effective energy of a magnetic flux tube of the form
of the Eq.(3) tends (for $d \gg 1/\sqrt{B_{m}}$) to the result calculated from the
Schwinger formula for a homogeneous magnetic field. One may think, that this is
unexpected, as the spectrum of a magnetic flux tube is continuous (with zero modes if
$\phi>1$) whereas the spectrum of a homogeneous magnetic field, has a different
structure- it is discrete and consists of the well known Landau levels. According to
the appendix C these two spectra are related in the limit of large $d$. Indeed, if in
the analysis of the appendix C, instead of the Gaussian magnetic field of the Eq. (2),
we have used the magnetic field of Eq. (3) we would find metastable states with
energies almost identical with those of Landau levels (we have checked this
numerically). Thus, the continuous spectrum of the magnetic field of Eq. (3)
approaches, for $d\gg 1/\sqrt{B_{m}}$, to the discrete spectrum of a homogeneous
magnetic field, in a straightforward way: the each metastable energy level approaches
a correspondingly Landau level and at the same time more metastable states are being
added at the high energy end.

\section{APPENDIX: Numerical calculation of the phase
shifts}

In this appendix we describe two ways for the numerical calculation of the phase
shifts.

We will adapt the method which was presented by E. Farhy $\mathit{et\: al}$ in
\cite{13,ph} to our problem. We consider two linearly independent solutions,
$u_{k,l,s}^{(1)}(r),\:u_{k,l,s}^{(2)}(r)$ of the radial equation
\begin{eqnarray}
(-\frac{d^{2}}{dr^{2}}+\frac{l^{2}-\frac{1}{4}}{r^{2}}+v_{l,s}(r)-k^{2})\:u_{k,l,s}(r)=0
\end{eqnarray}
with asymptotic behavior $e^{\pm ikr}$ as $r$ tends to infinity.

We can put these solutions into the form
\begin{eqnarray}
u_{k,l,s}^{(1)}(r)&=&e^{i\beta_{l,s}(k,r)}\sqrt{r}H^{(1)}_{l}(kr)\\
u_{k,l,s}^{(2)}(r)&=&e^{i\beta_{l,s}^{\ast}(k,r)}\sqrt{r}H^{(2)}_{l}(kr)
\end{eqnarray}
where $H^{(1)}_{l}(x)$ and $H^{(2)}_{l}(x)$ are the Hankel functions of the first and
second kind respectively. The complex function $\beta_{l,s}(k,r)$ tends to zero as $r$
tends to infinity.

 The scattering solutions $ u_{k,l,s}(r)$ of the radial equation (62)
satisfy the following boundary conditions:
\begin{eqnarray}
u_{k,l,s}(0)&=&0\\ u_{k,l,s}(r\longrightarrow+\infty)&\sim&\cos(k
r-\frac{\pi}{4}-\frac{l\pi}{2}+\delta_{l,s}(k))
\end{eqnarray}
The solutions $ u_{k,l,s}(r)$ are expressed as a linear
combination of the solutions (63) and (64):
\begin{eqnarray}
u_{k,l,s}(r)=e^{i\delta_{l,s}}e^{i\beta_{l,s}}\sqrt{r}H^{(1)}_{l}(kr)
+e^{-i\delta_{l,s}}e^{-i\beta^{\ast}_{l,s}}\sqrt{r}H^{(2)}_{l}(kr)
\end{eqnarray}
Imposing the boundary condition (65) to the solution (67) we
obtain
\begin{equation}
\delta_{l,s}(k)=-Re\:\beta_{l,s}(k,0)
\end{equation}
Substituting the solution (63) into the equation (62) we obtain the differential
equation:
\begin{equation}
i\beta_{l,s}''(k,r)+2ik\:q_{l}(kr)\beta_{l,s}'(k,r)-(\beta_{l,s}'(k,r))^{2}-v_{l,s}(r)=0
\end{equation}
where
\begin{equation}
q_{l}(x)=\frac{d}{dx}\Bigl\{ln(\sqrt{x}H^{(1)}_{l}(x))\Bigl\}
\end{equation}
The functions $\beta_{l,s}(k,r)$ should satisfy the following conditions:
\begin{eqnarray}
\beta_{l,s}(k,+\infty)&=&0 \\ \beta_{l,s}'(k,+\infty)&=&0
\end{eqnarray}
Because the potential tends to zero very slowly $O(1/r^{2})$, we
begin the numerical integration of the differential equation (69)
from a large number $r_{max}$ towards zero, with the conditions:
\begin{eqnarray}
\beta_{l,s}(k,r_{max})&=&(-2l\phi+\phi^{2})\frac{1}{2k\:r_{max}}
\\ \beta_{l,s}'(k,r_{max})&=&-(-2l\phi+\phi^{2})\frac{1}{2k\:r_{max}^{2}}
\end{eqnarray}
The above asymptotic behavior of the functions $\beta_{l,s}(k,r)$ can be obtained from
the equation (69).

Another way to compute the phase shifts is by solving the differential equation
\begin{equation}
\frac{d\del}{dr}=-\frac{\pi}{2}r\;v_{l,s}(r)[J_{l}(kr)\cos\del-N_{l}(kr)\sin\del]^{2}
\end{equation}
with the boundary condition $\delta_{l,s}(k,0)=0$ where $J_{l}(x)$ and $N_{l}(x)$ are
the Bessel and Neumann functions respectively. The phase shift is given by the limit
\begin{equation}
\delta_{l,s}(k)=\lim_{r\rightarrow+\infty}\delta_{l,s}(k,r)
\end{equation}

This method was formulated by F. Calogero in reference \cite{14} for the
$3$-dimensional case. It was obtained for two dimensions in reference \cite{15}.

For the numerical computations of the phase shifts we have used mainly the
differential equation (69) because it gives results faster and more accurate than the
equation (75). However, this differential equation is stiff for large values of $\phi$
and small values of $l$. In that region of $\phi$ and $l$ we have used the Eq. (75).

We also compared our numerical results with those of WKB approximation (22) and we
found good agreement for large $l$ or large $k$.

\end{document}